\documentclass[comsoc, journal]{IEEEtran}
\usepackage[utf8]{inputenc}
\usepackage[T1]{fontenc}

\usepackage{todonotes}
\usepackage{graphicx}
\usepackage{subfig}
\usepackage{float}

\usepackage[hyphens]{url}
 \usepackage{hyperref}

\usepackage{tabularx}

\usepackage[noadjust]{cite}

\usepackage{array, makecell, multirow, booktabs}
\newcolumntype{P}[1]{>{\centering\arraybackslash}p{#1}}
\setcellgapes{3pt}

\newcommand{\eg}{\textit{e.g.,~}}
\newcommand{\ie}{\textit{i.e.,~}}

\usepackage{color, xcolor}

\usepackage{xstring}
\usepackage{multirow}
\usepackage{subfig}
\usepackage{color, soul}
\usepackage{xspace}
\usepackage{caption}

\IEEEpubid{\begin{minipage}{\textwidth}\ \\[32pt] \centering
  \footnote {© 2021 IEEE. Personal use of this material is permitted. Permission from IEEE must be obtained for all other uses, in any current or future media, including reprinting/republishing this material for advertising or promotional purposes, creating new collective works, for resale or redistribution to servers or lists, or reuse of any copyrighted component of this work in other works.}
\end{minipage}} 

\title{Networking and Computing in Biomechanical Research: Challenges and Directions}

\author{
	Spyridon Mastorakis, Andreas Skiadopoulos, Susmit Shannigrahi, \\ Aaron Likens, Boubakr Nour, and Nicholas Stergiou
	
	\thanks{S. Mastorakis (corresponding author), A. Skiadopoulos, A. Likens and N. Stergiou are with the University of Nebraska at Omaha, US.}
	
	\thanks{S. Shannigrahi is with the Tennessee Tech University, US.}
	
	\thanks{B. Nour is with the Beijing Institute of Technology, China.}
}

\IEEEpeerreviewmaketitle

\begin{document}
\maketitle

\begin{abstract}
Biomechanics is a scientific discipline that studies the forces acting on a body and the effects they produce. In this paper, we bring together biomechanists and networking researchers to shed light into how research efforts in biomechanics, primarily related to the study of the human body, can be facilitated through networking and computing technologies, such as edge and cloud computing, Software Defined Networking, and Information-Centric Networking. We first present challenges related to networking and computing that biomechanists face today and we then describe how networking and computing technologies can address them. Finally, we identify directions for future networking research with a focus on biomechanics to facilitate and encourage interdisciplinary collaborations between biomechanists and networking researchers.
\end{abstract}

\section{Introduction}
\label{sec:intro}
Biomechanics aims to study, understand, predict, and explain the effects of forces and torques that act upon living biological systems. It has been defined as the ``study of the forces that act on a body and the effects they produce''~\cite{stergiou_biomechanics_2020}. As a scientific discipline, biomechanical methodologies are used in human movement research to explain the structural and functional mechanisms underlying movement performance.

Research activities in human movement biomechanics rely heavily on clinical experiments, the collection of data from human subjects, and the statistical analysis of collected data. However, such activities are often constrained due to the limited connectivity capabilities of the used equipment and the lack of synergy with networking and computing technologies. For example, commercial Motion Capture (MoCap) systems have large-scale bandwidth requirements, while they often utilize proprietary wireless technologies and dedicated network hardware to transfer captured data for processing, limiting flexibility and interoperability.
In this paper, we bring biomechanists and networking researchers together to demonstrate how biomechanical research can be enhanced through prominent networking and computing technologies. Through this paper, we aim to motivate the exploration of synergies among biomechanics, networking, and computing. 

The contribution of our work is two-fold: (i) we present the characteristics and requirements of biomechanical data collection and analysis, as well as the challenges in biomechanical research related to networking and computing; and (ii) we describe how technologies, such as edge computing~\cite{shi2016edge}, Information-Centric Networking (ICN)~\cite{afanasyev2018brief}, and Software-Defined Networking (SDN)~\cite{bannour2018distributed}, can contribute to resolving the presented challenges and we identify directions for future research with a focus on encouraging interdisciplinary collaborations between networking researchers and biomechanists.

Our work is organized as follows: in Section~\ref{sec:background}, we provide a brief background and the motivation for focusing on biomechanics as our use case. In Section~\ref{sec:challenges}, we present challenges of biomechanical research related to networking and computing, and, in Section~\ref{sec:networking}, we discuss how networking and computing can enable biomechanical research. In Section~\ref{sec:discussion}, we identify directions for future networking research with a focus on biomechanics, and, in Section~\ref{sec:conclusion}, we conclude our work.

\section{Background and Motivation}
\label{sec:background}

In this section, we give a background on biomechanics, and computing and networking technologies. 
We also motivate our focus on biomechanics as our use case and the characteristics of this use case.


\subsection{Human Movement Biomechanics}

Biomechanics is a field that uses the knowledge of mechanics to study the action of forces and their effects on living biological systems in macro- and micro-scale. 
Biomechanical research witnessed an explosion during the 20th century fueled by methodological and technological advancements. 
The technological progression 
has led to the development of compact sensors that reshaped human movement research by making it possible to collect time-dependent biomechanical data under complex tasks outside of laboratories.



Biomechanics sits at the intersection of several disciplines, including biology, physiology, anatomy, physics, mathematics, and chemistry. There is a broad spectrum of research areas in biomechanics that is intertwined with research in other fields, such as engineering and exercise science. In human movement research, the primary objective is to analyze the kinetics of underlying movement. Human movement kinetics investigates the cause-effect relationships between muscle forces and the time course of motion~\cite{stergiou_biomechanics_2020}. 
Biomechanical analysis describes movement using experimentally obtained kinetics data. 
Biomechanical analysis using musculoskeletal models estimates the muscle forces during the time course of motion.
Musculoskeletal biomechanics provides information about how muscle forces are coordinated to perform a specific task. Consequently, musculoskeletal analysis offers the possibility to understand how altered muscle coordination patterns affect movement performance in clinical populations. These populations include a broad spectrum of pathology, ranging from physiological and neurological decline to diabetes, stroke, and cerebral palsy.
To translate musculoskeletal biomechanics from laboratory to clinical settings, musculoskeletal models should match the patient’s conditions. However, such ``personalization'' needs extra computing time. 
In a similar fashion, musculoskeletal biomechanics may be used on the field using portable equipment to collect data~\cite{Karatsidis_2019}. 
Musculoskeletal biomechanics can contribute to domains such as medicine and rehabilitation, occupational safety and ergonomics, sports, automobile and aerospace design,
industrial engineering, and robotics.

\subsection {Networking and Computing Technologies}
As a prominent candidate for future Internet, ICN~\cite{afanasyev2018brief} makes data the focus of the communication context, decoupling it from its production or hosting location(s). ICN uses application-defined naming directly at the network layer for communication purposes. To this end, consumer applications request content by name--for example, consumer applications can retrieve the latest news headlines from BBC by sending a request for data with a name ``/bbc/latest-news/headlines''. SDN~\cite{bannour2018distributed} has been proposed to decouple the forwarding from the control plane making control plane programmable and logically centralized. The network intelligence is realized through one or more SDN controllers, which instruct network elements on how to handle different traffic types.



In mid-2000, cloud computing started gaining traction as a paradigm that offers computing resources without the need for active management by users. 
Cloud typically incurs long network delays, which cannot be tolerated by applications, such as Augmented Reality (AR), that require low user-perceived latency. In combination with the massive data amounts generated by Internet-of-Things (IoT), edge computing~\cite{shi2016edge} has emerged as a paradigm that pushes data processing and storage close to users. Finally, next-generation mobile technologies, such as 5G, are considered as enablers of the edge computing deployment by offering low-latency access to computing resources for billions of devices~\cite{parvez2018survey}, but have yet to be explored in biomechanics.


\subsection {Motivation}
\label{subsec:motivation}

\begin{figure}[tbp]
    \centering
    \includegraphics[width=0.40\linewidth]{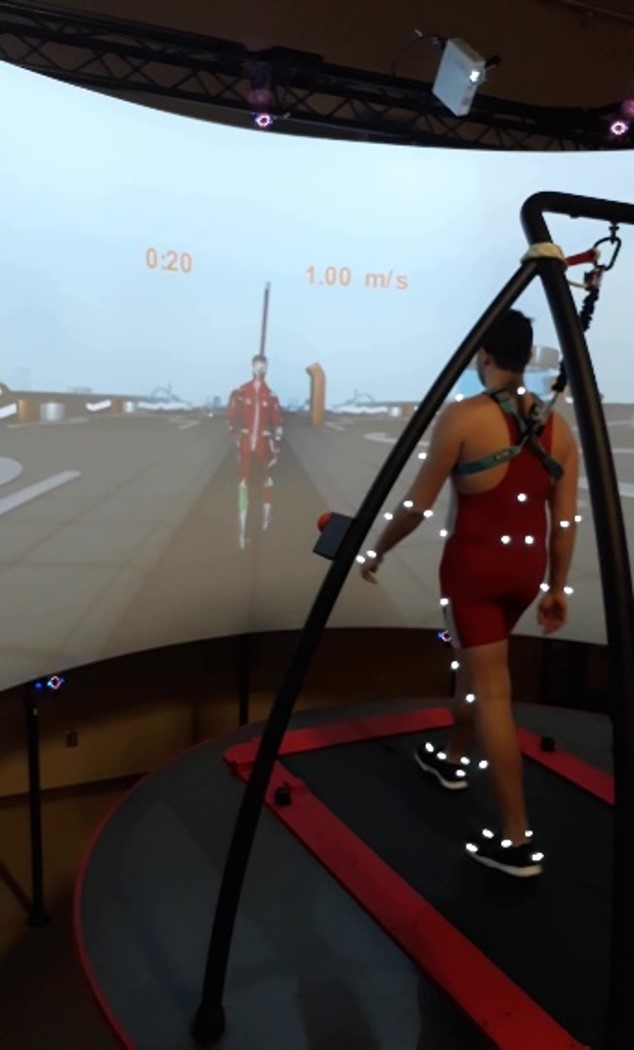}
    \vspace{-0.1cm}
    \caption{\small Biomechanical data collection in a laboratory through experiments that involve walking/running on a treadmill.} 
    \label{fig:lab}
    \vspace{-0.5cm}
\end{figure}

Table~\ref{table:motivation} presents studies on human movement that involve 
biomechanical data collection, analysis, and feedback 
in laboratory (indoor) spaces. Traditionally, movement is captured by 
video-based MoCap systems. A typical setup may include up to 12 cameras, which record reflections from markers placed on a human subject's body (Figure~\ref{fig:lab}). Such systems have large-scale bandwidth requirements often exceeding 100MB/s (800Mbps) per camera~\cite{vicon}. These studies can be extended for the collection of biomechanical data outdoors based on MoCap systems with wireless inertial measurement units~\cite{Karatsidis_2019}. 
Each unit may reach a data rate of up to 4Mbps, while commercial MoCap systems may include 17-32 such units. Considering that electromyography (EMG) data for muscle activity and pressure insoles to reconstruct Ground Reaction Forces (GRF) may also be needed, the overall bandwidth requirements may range between 72Mbps and 132Mbps per subject.

\begin{table}
\centering
\vspace{-0.2cm}
\caption{\small{Representative studies of biomechanical data collection and analysis.}}
\label{table:motivation}
\begin{tabularx}{0.45\textwidth}{|p{1.8cm}|p{1.45cm}|p{1.45cm}|p{1.8cm}|}
\hline Reference
                    & \cite{pizzolato2017biofeedback} & \cite{pizzolato2017real} & \cite{van2013real}   \\ \hline
Data sampling rate:     &           &           &  \\ 
          - MoCap    & 200 Hz    & 200 Hz    & 100 Hz  \\ 
          - GRF      & 1000 Hz   & 1000 Hz   & 1000 Hz   \\ 
          - EMG      & 2000 Hz   & N/A         &  N/A  \\ \hline
Analysis outcomes (feedback) & Knee joint load & Joint angles and moments & Joint angles and moments, muscle forces, knee joint load    \\ \hline
Musculoskeletal model &  Personalized & Personalized & Generic  \\ \hline
Computing time (ms) &  61 & 11 & 6.72  \\ \hline
Total data rate (Gbps) & $\sim$9.6 & $\sim$9.6 & $\sim$12.8 \\ \hline
Experimental context & Indoors/Can be extended for outdoors & Indoors/Can be extended for outdoors & Indoors/Can be extended for outdoors \\ \hline
\end{tabularx}
\vspace{-0.5cm}
\end{table}


A characteristic and challenge of biomechanical research is its potentially high data sampling frequency, ranging from 100 to 2,000 samples per second. This frequency depends on the type of the collected data (\eg MoCap data for 3D marker coordinates, EMG and GRF data). This demonstrates the need to collect 
different data types in order to perform meaningful 
analysis at the muscle level, while the collected data and analysis outcomes (\eg kinematics, kinetics, and muscle forces) need to be highly precise. The time required for analysis (computing) of the collected data ranges from roughly 7ms to 61ms. The lower bound can be achieved through generic musculoskeletal models and state-of-the-art equipment and software, such as the computer assisted rehabilitation environment, which may cost more than a million dollars. Personalized musculoskeletal models and commodity software and equipment can achieve the upper bound. 

Due to the human neurophysiological perception, the maximum response time that biomechanical analysis and feedback can tolerate is 75ms~\cite{kannape2013self}. This requirement signifies that the tolerable latency for biomechanical analysis may be dominated by computing, leaving little flexibility for the transmission of data to computing resources. However, the data transmission time is a crucial factor when it comes to expanding biomechanics to outdoor environments, which offer insights to habitual behaviors of users in daily living contexts.
In such environments, 
collected data needs to be 
transmitted for biomechanical analysis to remote computing resources. This highlights the need for the integration of cutting-edge networking and computing technologies in biomechanics, including core networking and wireless technologies previously discussed in a broader healthcare context~\cite{cisotto2020requirements}.

The TCP/IP architecture may not be the most suitable option for biomechanics, since it relies on end-to-end connections, which are sensitive to loss, thus not fully utilizing high bandwidth links. The biomechanics community requires high throughput in the presence of loss and low latency, a requirement that can be achieved by technologies such as SDN, ICN, and 5G coupled with cloud and edge computing.


\section{Challenges of Biomechanical Research Related to Networking and Computing} 
\label{sec:challenges}

In this section, we highlight networking and computing challenges that biomechanists face. Table~\ref{Table:challenges} presents a summary and categorization of these challenges.

\begin{table*}[!t]
\centering
\caption{Challenges related to computing and networking in biomechanics}.
\vspace{-0.2cm}
\resizebox{0.86\textwidth}{!}{
\begin{tabular}{lll}
\toprule
\multicolumn{1}{c}{\textbf{Use-case}}                                                & \multicolumn{2}{c}{\textbf{Challenges}}                                                                                                                                                    \\ \midrule
\multicolumn{1}{c}{}                                                                 & \multicolumn{1}{c}{\textbf{Computing}}                                                                                                                                         & \multicolumn{1}{c}{\textbf{Networking}}                                                                                                                      \\ \midrule
\begin{tabular}[c]{@{}l@{}} Laboratory-based\\ {biomechanical experiments}\end{tabular}   & \begin{tabular}[c]{@{}l@{}}$\bullet$ Complex mathematical models for data analysis \\ $\bullet$ Correlation of biomechanical data with \\ previously collected data to identify disease\end{tabular} & \begin{tabular}[c]{@{}l@{}}$\bullet$ Identification of continuous data streams with  \\ different frequencies from different equipment \end{tabular}                                   \\ \midrule
\begin{tabular}[c]{@{}l@{}} Low-latency 3D motion \\ tracking in the wild\end{tabular} & \begin{tabular}[c]{@{}l@{}} $\bullet$ Biomechanical analysis of (potentially \\ high-frequency) data as a service close to users\end{tabular}                                                                                                           & \begin{tabular}[c]{@{}l@{}}$\bullet$ Limited/proprietary wireless connectivity \\ capabilities of MoCap systems\end{tabular}                                            \\ \midrule
\begin{tabular}[c]{@{}l@{}}Data management, replication,\\ and security\end{tabular}   & \begin{tabular}[c]{@{}l@{}}$\bullet$ Efficient management of biomechanical data \end{tabular} & \begin{tabular}[c]{@{}l@{}}$\bullet$ Seamless data replication for long-term storage\\ $\bullet$ Biomechanical data security at rest and in transit\end{tabular}           \\ \midrule
AR human gait metronomes                                                             & \begin{tabular}[c]{@{}l@{}}$\bullet$ Low-latency processing of biomechanical\\ data and AR-based, ``gamified'' stimuli\end{tabular}                                                      & \begin{tabular}[c]{@{}l@{}}$\bullet$ ``Aggregated'' offloading of different data types\\ for meaningful biomechanical feedback\end{tabular}                                                    \\ \midrule
\begin{tabular}[c]{@{}l@{}}Physical activity tracking\end{tabular}                                                        & \begin{tabular}[c]{@{}l@{}}$\bullet$ Low-latency biomechanical feedback \end{tabular}                             & \begin{tabular}[c]{@{}l@{}} $\bullet$ ``Aggregated'' data offloading for meaningful \\  biomechanical feedback\end{tabular} \\ \bottomrule
\end{tabular}
}
\label{Table:challenges}
\vspace{-0.6cm}
\end{table*}



\textbf{Laboratory-based experiments:} In laboratories (Figure~\ref{fig:lab}), biomechanists conduct experiments by collecting continuous streams of different data types (\eg 3D linear and angular kinematics and their derivatives, external forces applied to body segments) through various instrumentation systems (\eg force plates, goniometers, accelerometers, markers on human body). Physiological (\eg heart and respiratory rates) and biological (\eg myoelectric signals) data may be simultaneously collected. The collected data is processed based on complex mathematical models.
The collected data may be sampled with high frequencies and require high precision, while the different data types (or data produced by different instruments) may have different sampling frequencies.
To enable the analysis of data of different types and frequencies, we need to easily identify the source of each data piece (\ie which instrument produced each data piece). Given that biomechanical analysis is performed on time-series of collected data and the analysis outcomes are represented as time-series, we also need to identify when each data piece was generated. This calls for flexible and accurate mechanisms of data identification/naming and timestamping.

Another challenge is associated with correlating collected data with previously collected data from subjects of different profiles (\eg age, sex, height) and disease (\eg multiple sclerosis, Parkinson's). This enables the classification of subjects based on whether their gait resembles the gait of healthy, young adults or older adults with a certain disease, acting as an initial disease indication. This process should happen online during biomechanical experiments, so that an outcome is produced by the end of the experiments. For example, if subjects are instructed to walk on a treadmill for a few minutes, the collected data samples need to be transferred to the classification algorithm as soon as they are produced, so that the outcome is ready by the end of the experiment.

\begin{figure}[tb]
    \centering
    \includegraphics[width=0.40\linewidth]{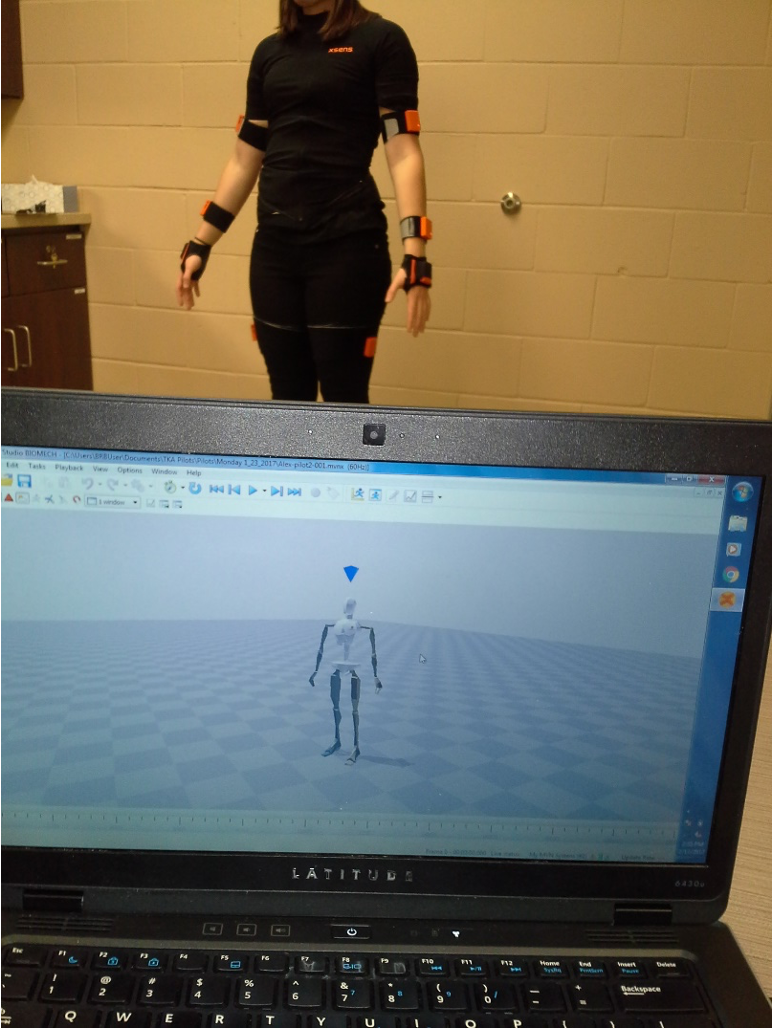}
    \caption{\small A subject wearing an Xsens suit~\cite{paulich2018xsens} along with the subject's 3D silhouette displayed on a laptop.} 
    \label{fig:xsens}
    \vspace{-0.6cm}
\end{figure}

\textbf{Low-latency 3D motion tracking in the wild:} Video-based MoCap systems are used to track movements in various settings. However, the covering space of such systems is fixed and they can be mostly used within laboratories. Data collection outside of laboratories is needed, since it provides insights to habitual behaviors (\eg walking on a university campus), which may not be possible in laboratories. To enable low-latency 3D motion tracking outside of laboratories, MoCap systems with inertial measurement units and wireless communication capabilities have become available (Figure~\ref{fig:xsens}). 

Such commercial MoCap systems offer limited connectivity options with Bluetooth-based, dedicated WiFi-based, or proprietary wireless solutions being common options. However, we would like to offer biomechanical analysis as a service through standardized wireless protocols that can satisfy the large-scale bandwidth requirements of MoCap systems. Existing solutions restrict biomechanists, since they follow one of the following approaches: (i) store the collected data on a USB drive mounted to the MoCap system, so that the data can be analyzed after the experiment; (ii) utilize the low range and bandwidth Bluetooth connection to transmit data to a laptop or mobile phone for analysis or visualization respectively, limiting the scope and feasibility of experiments; or (iii) use proprietary wireless protocols and/or a dedicated wireless Access Point (AP) connected to a laptop to transmit data from the MoCap system to the laptop making troubleshooting difficult, and limiting interoperability and flexibility.


\textbf{Data management, replication, and security:} The most common practice for biomechanical data exchanges among scientists is to store the data collected during an experiment as separate files (\eg ASCII, text, 3D biomechanics data standard) for each subject. These files may be encrypted and password protected, residing on servers that comply with national and institutional security and privacy requirements. However, there are no systematic ways to: (i) organize and manage the collected data; (ii) replicate the data across different servers for long-term storage to deal with hardware failures; and (iii) secure the data at rest and in transit. 

\textbf{Metronomes for gait adjustments:} Auditory and visual metronomes have been used in biomechanics to study the effects of sound and image patterns on gait~\cite{stergiou_biomechanics_2020}, 
manipulating gait characteristics and synchronizing gait to the metronomes. For example, subjects listening to the auditory stimulus of a metronome may be instructed to walk on a flat surface to determine whether they can synchronize their gait to the metronome beat. To capture gait data, such as stride length and intervals (the time between two consecutive heel contacts of the same leg), cameras along with foot switches or wireless sensors may be used. 

A direction that can enable tremendous biomechanical research is turning AR headsets into AR gait metronomes and enabling their operation indoors and outdoors. Such metronomes will augment the perception of the world around subjects, offering enriched, interactive, and ``gamified'' stimuli. For example, subjects may be instructed to synchronize their steps with visual targets (\eg step on 3D bulls-eye objects visualized on the floor by the AR headset). Biomechanical (\eg gait stride length and interval) and AR data (\eg scenes including the visual targets) needs to be offloaded to available computing resources and be processed with low latency, so that the AR stimulus can adapt based on the steps of the subjects under different conditions (\eg slow pace walking, running). For example, if subjects based on their stride length and the AR scenes are not detected to step on the visual targets, they should be reminded to do so (\eg through messages displayed on the headset's screen). 
A challenge to consider is that biomechanical and AR data needs to be offloaded and processed by the same edge server, so that meaningful feedback can be provided to subjects.

\textbf{Physical activity tracking:} A focus area of biomechanics has to do with exercise training effects and adaptation, exploring non-obstructive tracking systems that monitor biomechanical variables (\eg posture, forces, acceleration) for low-latency feedback to individuals. Commonly used devices may include smart wristbands and other wearable devices, or smart sport equipment (\eg wheelchairs or tennis racquets). Enabling such devices to provide low-latency feedback to individuals about their performance are challenges that expand the horizons of biomechanics, if resolved. A particular challenge to consider is the need to offload data from different devices to the same edge server for meaningful feedback. For example, to provide feedback about the physical activity of a wheelchair user, data from sensors placed on the wheelchair (to capture its speed and acceleration) and wearable sensors placed on the user's body needs to be aggregated and processed together.

\section{How Can Networking and Computing Enable Biomechanical Research?}
\label{sec:networking}

\begin{figure}[tbp]
    \centering
    \includegraphics[width=0.95\linewidth]{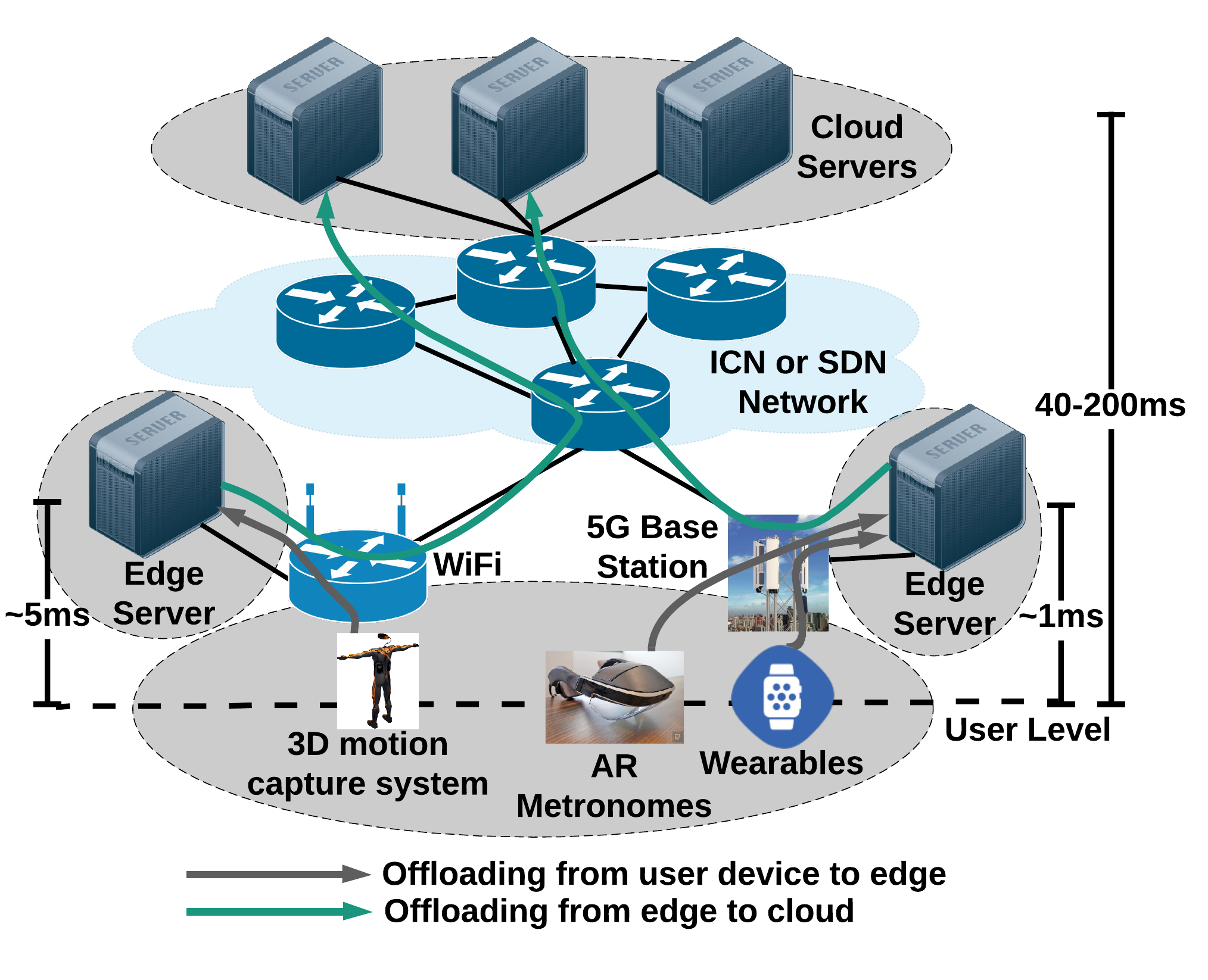}
    \vspace{-0.5cm}
    \caption{\small Enabling biomechanical research through prominent networking and computing technologies. Network delays from devices to edge and cloud servers are outlined.}
    \label{fig:networking}
    \vspace{-0.6cm}
\end{figure}

In this section, we discuss how networking and computing can address the challenges of Section~\ref{sec:challenges}. We present a summary of our discussion in Table~\ref{table:summary} and an example architecture in Figure~\ref{fig:networking}. This architecture consists of: (i) devices that offload biomechanical data and computation to edge servers; (ii) wireless and mobile technologies that provide connectivity to edge servers; (iii) edge servers with computing and storage resources that offer low-latency data processing, while they also offload raw and processed data to cloud servers for long-term statistical analysis and storage; (iv) an ICN- or SDN-capable network, providing connectivity across the edge and towards cloud servers; and (v) cloud servers with computing and storage resources.

\begin{table*}[!t]
\centering
\caption{Networking and computing paradigms and their benefits for biomechanical research.}
\vspace{-0.2cm}
\resizebox{0.86\textwidth}{!}{
\begin{tabular}{ccl}
\toprule
\textbf{Paradigm} & 
\textbf{Focus} &
\multicolumn{1}{c}{\textbf{Benefits}} \\ \midrule
\begin{tabular}[c]{@{}l@{}} Edge \\ Computing\end{tabular} & 
Computing &
\begin{tabular}[c]{@{}l@{}} $\bullet$ Low-latency, high precision biomechanical data processing\\ $\bullet$ Low-latency biomechanical feedback \\ $\bullet$ Low-latency machine learning to predict disease\end{tabular}                                            \\ \midrule
\begin{tabular}[c]{@{}l@{}} Cloud \\ Computing \end{tabular}                                               & Computing      & \begin{tabular}[c]{@{}l@{}} $\bullet$ Archiving of biomechanical data with different sampling frequencies\\ $\bullet$ Latency-tolerant operations\end{tabular}                                                                                                \\ \midrule
ICN                                                      & Networking     & \begin{tabular}[c]{@{}l@{}}$\bullet$ Seamless biomechanical data replication and increased data security\\ $\bullet$ Flexible naming for biomechanical data produced by different equipment \\ 
$\bullet$ ``Aggregated'' offloading of different data types to the edge\end{tabular} \\
\midrule
SDN                                                      & Networking     & \begin{tabular}[c]{@{}l@{}}$\bullet$ Facilitates (combined with NFV) the deployment of biomechanical analysis functions \\$\bullet$ Accurate time synchronization through the controller\\ $\bullet$ ``Aggregated'' offloading of different data types to the edge\end{tabular}         \\ \midrule
\begin{tabular}[c]{@{}l@{}}5G and \\ beyond\end{tabular} & Networking     & \begin{tabular}[c]{@{}l@{}}$\bullet$ High-bandwidth and low-latency connectivity for MoCap systems \\ $\bullet$ Enabling biomechanical data collection outdoors\\   $\bullet$ Slicing for biomechanical data with different quality of service requirements\end{tabular}      \\ \bottomrule

\end{tabular}
}
\label{table:summary}
\vspace{-0.6cm}
\end{table*}

\textbf{Cloud and edge computing:} Cloud and edge computing facilitate the processing of biomechanical data based on complex, compute-intensive mathematical models. These models can be deployed at the edge to offer response times of less than 75ms.
Processed data can be archived and non time-sensitive operations (\eg long-term statistical processing) can be executed on the cloud based on hybrid edge-cloud computing models. To verify the feasibility of edge computing for biomechanics, we deployed a commodity Linux server connected to a 5GHz WiFi AP. The latency from devices to the server within the same WiFi network is 3-6ms, being able to achieve response times of less than 75ms based on the computing times of Table~\ref{table:motivation}. On the other hand, access to cloud resources may require 40ms-200ms depending on the distance between a client and a cloud server, thus providing the required response times through the cloud may not be feasible.


Machine learning at the edge can identify patterns and disease based on collected gait data in an online fashion with low latency. Supervised learning models can correlate gait data collected from a new individual with a database of collected, labelled gait data from individuals of different ages and with different diseases. Given that biomechanical studies may include small numbers of human subjects, in cases where labelled data is limited or non-existent (\eg for individuals with a rare disease), solutions based on semi-supervised or reinforcement learning may be helpful. 

\textbf{ICN}: Deploying large numbers of sensors and pieces of equipment based on an IP-based point-to-point communication infrastructure may be cumbersome to maintain and scale. On the contrary, ICN 
is well-suited for data addressing, replication, and distribution through the use of application-defined naming at the network layer. Flexible namespaces can be designed to identify and retrieve data generated by different equipment, which may have different data generation frequencies. For example, a namespace ``/sensors/<subject-id>/<data-type>/<sample-number>'' facilitates the retrieval of samples of continuous data streams from sensors of different types placed on a subject's body without complex addressing. ICN naming also facilitates the adaptive offloading of data and computational tasks. This ensures, for example, that data from all the sensors placed on a subject is offloaded to the same among the available edge servers, so that this data is aggregated and processed together. This further facilitates AR metronomes, where biomechanical and AR data needs to be processed together. ICN has been proven substantially more resilient than IP for data distribution under dynamic connectivity and wireless links, achieving 15-33\% lower data distribution times than IP-based solutions~\cite{mastorakis2020dapes}.


Furthermore, ICN offers mechanisms for automated trust management and verification based on data names. Biomechanical data may be privacy sensitive, thus it can benefit from ICN's data-centric security, since data is signed and can also be encrypted by its producer, while it can be decrypted only by entities with the proper key(s). This is a contrast to the channel-based TLS/SSL security typically used in TCP/IP, especially under mobility where secure channels may break.


\textbf{SDN:} SDN contributes to efficient network management, complementing the computing advancements that enable biomechanical research. Combined with Network Function Virtualization (NFV), SDN enables the deployment of biomechanical analysis services at the edge or cloud regardless of the underlying hardware. Hybrid SDN-ICN architectures make use of ``the best of both worlds'' by leveraging: (i) the stateful name-based ICN forwarding plane; and (ii) the (logically) centralized SDN control plane to orchestrate the ICN forwarding plane. 

Such architectures enable the aggregation of different data types and their offloading to the same edge server for meaningful processing and feedback. SDN can effectively create forwarding rules and install them on network elements, so that the network can forward different data types, which need to be processed together, to the same edge server. In this case, the aggregation of different data types will happen at the edge server before processing. Combined with NFV, data plane virtual network functions can be offloaded to programmable switches, which in turn enable in-network computing paradigms. In this case, the aggregation of different data types can happen in the network, and the aggregated data will be subsequently forwarded to the same edge server for processing. Finally, SDN can: (i) handle the mobility of subjects during biomechanical experiments; and (ii) re-route offloaded data through non-congested paths and to edge servers with available resources in cases of network congestion and overloaded servers respectively, in addition to relocating processing functions from overloaded servers to servers with available resources.


The SDN controller can act as a central point for time synchronization between different biomechanical equipment, improving the accuracy of the timestamps of generated data. Evaluation results have shown that SDN enables accurate time synchronization with negligible errors in the order of 10-44ns~\cite{li2020sdn}. An SDN controller constitutes a single point of failure, a weakness that can be mitigated through distributed control planes falling under two major categories~\cite{bannour2018distributed}: (i) maintaining multiple replicas of an SDN controller in a flat fashion; and (ii) creating a hierarchy of SDN controllers. In category (i), a primary controller is elected, while replicas maintain a copy of the primary's state. In category (ii), each hierarchy layer includes controllers with different responsibilities, so that frequent local events can be handled by the bottom layer, while non-local events in need of a network-wide view can be handled by higher layers.


\textbf{5G and beyond:} 5G and beyond technologies can enable low-latency and high-bandwidth access to edge computing resources increasing the capabilities of AR metronomes and empowering MoCap systems to operate without proprietary wireless protocols or dedicated APs. With the potential to achieve data rates up to 20Gbps and end-to-end latency in the order of 1ms~\cite{parvez2018survey}, 5G can enable experimental setups for biomechanical data collection outside of laboratories. 
Such large-scale data rates are needed by MoCap systems, while low network latency is also needed to provide the required response times for biomechanical analysis and feedback. 
The network slicing capabilities of 5G can further facilitate the delivery of biomechanical data with different characteristics. 
Biomechanical data may have different sampling frequencies, thus high and low frequency data may have different slices. In addition, sensitive biomechanical data (\eg containing identifiable information), data with strict latency requirements (\eg AR data), and mission critical data (\eg heart rate readings) may have different slices. In such cases, each slice will have custom parameters to support different requirements (\eg security and privacy, latency and/or bandwidth requirements, guaranteed delivery). 



\section{Future Research Directions}
\label{sec:discussion}


In this section, we identify research directions for the networking community to enable biomechanical research. 

\textbf {Discovery of resources under mobility:} We envision biomechanical data being collected while users move freely in daily living contexts outdoors. As users move, they access different edge networks, which may offer different computing functions and resources of different capabilities. To this end, mechanisms to dynamically discover resources and functions at the edge are needed. 
Such mechanisms can be explored in the context of SDN, where they are aided by the logically centralized control plane, and ICN, where they can be performed in a distributed manner from within the network. 

\textbf {Enhanced wireless capabilities of biomechanical equipment}: Enhancing MoCap systems and wearable devices with comprehensive wireless capabilities for the transmission of biomechanical data 
is vital. Such enhancements will enable the collection of biomechanical data outdoors with the minimum required equipment. The integration of MoCap systems with standardized technologies such as 5G for mobile communication and 802.11ac for WiFi communication should be investigated, since such systems require high-bandwidth and low-latency connections. 


\textbf {Securing biomechanical data:} 
To enhance security at rest, trusted hardware approaches, such as Intel SGX and ARM Trustzone, 
should be considered. Such approaches require a careful tradeoff assessment, since they may result in performance degradation as a consequence of increased security, which may be amplified due to the high sampling frequencies of biomechanical data. To secure data transmission, ICN offers a promising network architecture, however, issues related to trust management, certificate distribution and revocation, and name-based access control should be explored.


\textbf {Hybrid edge-cloud for data with high sampling frequencies:} In biomechanics, data is produced at the edge and needs to be eventually stored towards the network core. 
Given the high sampling frequencies of biomechanical data, the data can be first processed at the edge, so that low-latency feedback is provided to subjects. Subsequently, the raw and processed data will be aggregated at the edge and will be forwarded to the cloud for long-term storage and analysis. Mechanisms to aggregate data of different frequencies at the edge and transfer it to the cloud should be explored.


\textbf {Machine learning and artificial intelligence:} Machine learning and artificial intelligence at the edge can identify the gait characteristics of individuals based on collected gait data. Directions involving the improvement of machine learning models and mechanisms to securely transfer and de-identify the data that will be processed by such models deserve further investigation. Finally, a challenge to be tackled is related to designing learning approaches for cases where the available labelled data is limited or non-existent.

\section{Conclusion}
\label{sec:conclusion}

In this paper, we discussed how networking and computing can enable biomechanical research. First, we highlighted challenges related to networking and computing that hinder biomechanical research efforts. Subsequently, we discussed how prominent networking and computing technologies can enable biomechanical research  by addressing these challenges. 
Finally, we highlighted research directions to facilitate interdisciplinary collaborations between networking researchers and biomechanists. We hope that this paper will motivate networking researchers to explore synergies and applications of their research to biomechanics.

\section*{Acknowledgements}
This work is supported by the NIH (P20GM109090), NSF (CNS-2016714), and the Nebraska University Collaboration Initiative. 


\bibliographystyle{IEEEtran}

\section*{Biographies}

\vskip -5.0\baselineskip plus -2fil

\begin{IEEEbiographynophoto}{Spyridon Mastorakis}
(smastorakis@unomaha.edu) is an Assistant Professor in Computer Science at the University of Nebraska Omaha. He received his Ph.D. in Computer Science from UCLA in 2019. 
His research interests include network systems, edge computing, IoT, and security.
\end{IEEEbiographynophoto}

\vskip -4.0\baselineskip plus -2fil

\begin{IEEEbiographynophoto}{Andreas Skiadopoulos}
(askiadopoulos2@unomaha.edu) received his 
Ph.D. in Human Movement Science from the University of Extremadura in 2016. He is a Research Associate in Biomechanics, University of Nebraska Omaha. His interests include applied neuromechanics.
\end{IEEEbiographynophoto}

\vskip -4.0\baselineskip plus -2fil

\begin{IEEEbiographynophoto}{Susmit Shannigrahi}
(sshannigrahi@tntech.edu) is an Assistant Professor in Computer Science at Tennessee Tech. He received his Ph.D. from Colorado State University in 2019. His interests include next generation networking.
\end{IEEEbiographynophoto}

\vskip -4.0\baselineskip plus -2fil

\begin{IEEEbiographynophoto}{Aaron Likens}
(alikens@unomaha.edu) received his Ph.D. from the Arizona State University. He is an Assistant Professor in Biomechanics at the University of Nebraska Omaha. His interests include nonlinear dynamics. 
\end{IEEEbiographynophoto}

\vskip -4.0\baselineskip plus -2fil

\begin{IEEEbiographynophoto}{Boubakr Nour} (n.boubakr@bit.edu.cn) received his Ph.D. in Computer Science and Technology at the Beijing Institute of Technology. His interests include next-generation networking. 
\end{IEEEbiographynophoto}

\vskip -4.0\baselineskip plus -2fil

\begin{IEEEbiographynophoto}{Nicholas Stergiou} (nstergiou@unomaha.edu) is the Distinguished Community Research Chair in Biomechanics and the Director of the Center for Research in Human Movement Variability, University of Nebraska Omaha. 
His research focuses on understanding variability in human movement. 
\end{IEEEbiographynophoto}

\vskip -4.0\baselineskip plus -10fil

\end{document}